\documentclass[aps,prx,twocolumn,final,letterpaper]{revtex4-2}

\usepackage{appendix}
\usepackage{graphicx}   
\usepackage{import}
\usepackage{comment}
\usepackage{epstopdf}
\usepackage{bm}
\usepackage{amssymb}
\usepackage{quotes}
\usepackage{braket}
\usepackage{transparent}
\usepackage{dcolumn}% Align table columns on decimal point
\usepackage{multirow}
\usepackage{cancel} % for the command \cancel and \cancelto
\usepackage{mdframed}
\usepackage{color}
\usepackage{bm}
\usepackage{dsfont}
\usepackage{slashed}
\usepackage{enumitem}
\usepackage{amsmath} 
\begin{document}
\rmfamily

%-----TITLE-----
\title{Probing intensity noise in ultrafast pulses using the dispersive Fourier transform augmented by quantum sensitivity analysis}

%-----AUTHORS AND AFFILIATIONS-----
\author{Shiekh Zia Uddin$^{1,2,\dagger}$}
%\email{suddin@mit.edu}
\author{Sahil Pontula$^{1,2,3,\dagger}$}
%\email{spontula@mit.edu}
\author{Jiaxin Liu$^4$}
\author{Shutao Xu$^5$}
\author{Seou Choi$^{2,3}$}
\author{Michelle Y. Sander$^5$}
\author{Marin Solja\v{c}i\'{c}$^{1,3}$}
\affiliation{$^1$ Department of Physics, MIT, Cambridge, MA 02139, USA. \\
$^2$ Research Laboratory of Electronics, MIT, Cambridge, MA 02139,  USA.\\
$^3$ Department of Electrical Engineering and Computer Science, MIT 02139, Cambridge, MA, USA.\\
$^4$ Department of Physics, Imperial College London, South Kensington, London SW7 2BW, UK.\\
$^5$ Department of Electrical and Computer Engineering and BU Photonics Center, Boston University, Boston, MA 02215, USA.\\
$\dagger$ Denotes equal contribution. Email: suddin@mit.edu, spontula@mit.edu}

\begin{abstract} 

To reach the next frontier in multimode nonlinear optics, it is crucial to better understand the classical and quantum phenomena of systems with many interacting degrees of freedom -- both how they emerge and how they can be tailored to emerging applications, from multimode quantum light generation to optical computing. Soliton fission and Raman scattering comprise two such phenomena that are ideal testbeds for exploring multimode nonlinear optics, especially power-dependent physics. To fully capture the complexity of such processes, an experimental measurement technique capable of measuring shot-to-shot pulse variations is necessary. The dispersive Fourier transform (DFT) is the ideal technique to achieve this goal, using chromatic dispersion to temporally stretch an ultrafast pulse and map its spectrum onto a measurable temporal waveform. Here, we apply DFT to explore the power-dependent mean field and noise properties of soliton fission and Raman scattering. To explain quantum noise properties, the traditional approach is to perform several hundred stochastic simulations for computing statistics. In our work, we apply quantum sensitivity analysis (QSA) to compute the noise in any output observable based on fluctuations in the input pulse, all using a single backwards differentiation step. We find that the combination of DFT and QSA provides a powerful framework for understanding the quantum and classical properties of soliton fission and Raman scattering, and can be generalized to other multimode nonlinear phenomena.

%While DFT has been applied to study noise properties in nonlinear fibers, the power dependence of these properties and their correlation with classical processes such as soliton fission and Raman scattering have not been explored. We show how single- and two-mode intensity noise statistics show strong power dependence and support low-noise states, despite large amounts of noise on the input pulse. We complement our measurements with a quantum sensitivity analysis, exploring the dependence of fluctuations in the output spectrum on noise in the input spectrum. This theoretical analysis shows good agreement with the mean field and noise properties experimentally measured by DFT. Our results take the first steps towards fully characterizing the power-dependent mean field and quantum statistics in ultrafast multimode nonlinear systems.

\end{abstract}

\maketitle

\section{Introduction}
\label{sec:intro}

\begin{figure*}
    \centering
    \includegraphics[scale=0.7]{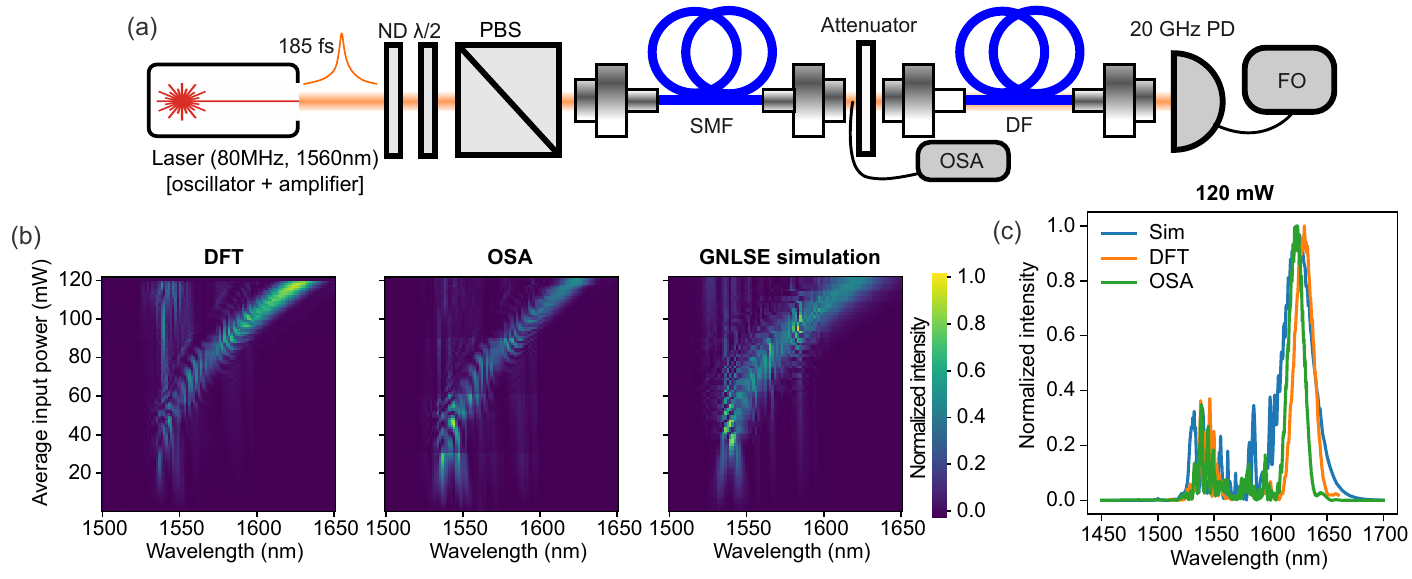}
    \caption{\textbf{Experimental setup and mean field intensity spectrum measurement.} (a) Experimental schematic for DFT measurement. (b) Intensity spectrum as a function of input pulse power for DFT, OSA, and GNLSE simulation. (c) Comparison of intensity spectra measured by DFT/OSA and calculated by GNLSE at input power of 120 mW. ND: neutral density filter, PBS: polarizing beam splitter, SMF: single mode fiber, OSA: optical spectrum analyzer, DF: dispersive fiber, PD: photodiode, FO: fast oscilloscope. In simulations, a fiber length $L=10$ m, Kerr nonlinearity $\gamma=10^{-3}$ 1/(W-m), and silica dispersion coefficients $\beta_2=-2.4\times 10^{-2}$ ps$^2$/m, $\beta_3=10^{-4}$ ps$^3$/m were used. Effects of Raman scattering (fractional Raman contribution $f_R=0.18$) were included.}
    \label{fig:1}
\end{figure*}

The intersection of multimode photonics and nonlinear optics has led to a burst of interest in developing multimode nonlinear systems with increasingly complex functionalities, ranging from optical computing and machine learning \cite{dang2024simple, oguz2024programming, yildirim2023nonlinear, rahmani2022learning, sunada2020using} to quantum light generation \cite{cristiani2022roadmap, wright2022nonlinear, yanagimoto2023mesoscopic, guidry2022quantum, guidry2023multimode, lustig2024emerging, kues2019quantum, roslund2014wavelength, ra2020non, presutti2024highly, yang2021squeezed}. These applications rely on phenomena enabled uniquely by having many interacting degrees of freedom. However, one of the key challenges inherent in these multimode optical systems is understanding their physical underpinnings, both from the perspective of describing how well-defined macroscopic physical phenomena arise from complex systems with many nonlinear interactions, and from the perspective of better controlling these systems for new applications.

Key insights have emerged from mean field theories such as optical thermodynamics \cite{wright2022physics, wu2019thermodynamic, makris2020statistical, ferraro2024calorimetry}, which abstract away individual microscopic nonlinear interactions and find simple laws governing macroscopic state variables that describe multimode nonlinear systems. Furthermore, insights from machine learning have provided ways to predict, control, and shape the mean field output of these systems \cite{rahmani2018multimode, teugin2020controlling, tzang2018adaptive, finot2024machine}.

The next frontier is understanding and controlling emergent quantum phenomena in multimode nonlinear systems. Exciting work in this direction has suggested ways to predict and generate entanglement, single and multimode squeezing, and other quantum states in nonlinear systems harboring multiple frequency modes \cite{reimer2016generation, uddin2023ab, pontula2024shaping, presutti2024highly}. It has become clear that a standard method to experimentally probe noise in highly multimoded systems is essential to both understand the noise properties of well-known classical nonlinear processes, as well as to engineer quantum statistics in new multimode quantum devices. Given that many multimode systems operate in the high-power, ultrafast regime to generate strong nonlinearity, shot-to-shot fluctuations and instabilities should be captured by a real-time noise measurement method.

To this end, the dispersive Fourier transform (DFT) technique achieves fast, continuous, single-shot measurements to detect rare and transient events \cite{goda2009theory, goda2013dispersive, godin2022recent}. In DFT, the spectrum of an ultrafast pulse is mapped through chromatic dispersion onto a temporal waveform that can be probed using a fast oscilloscope. DFT has been applied to study mean field soliton behavior \cite{wang2020recent, lapre2020dispersive} as well as spectral noise and correlations in supercontinuum generation \cite{wetzel2012real, godin2013real}. While nonlinear processes such as soliton fission and Raman scattering are known to be power dependent, their accompanying noise properties at variable power remain unexplored. 

In this work, we use the dispersive Fourier transform to probe the power dependent noise properties of an ultrafast pulse propagating through a nonlinear fiber. We calculate mean field spectra as well as one- and two-mode quantum noise properties, showing how low noise states are present despite large amounts of excess (classical) noise in the input pulse. The noise properties show sensitive power dependence and can be correlated with the onset of soliton fission and Raman scattering. We complement our experimental measurements with simulations based on quantum sensitivity analysis (QSA), a recently proposed method for calculating noise in an arbitrary output variable by using the adjoint method to calculate its ``sensitivity'' (Jacobian) with respect to input parameters \cite{uddin2023ab}. Through a QSA-augmented generalized nonlinear Schrodinger equation model, we explore how the intensity noise of the input pulse affects the noise properties of the output pulse, observing that some wavelengths in the output spectrum are insensitive to large amounts of added noise on the input spectrum at certain wavelengths. QSA also helps provide a physical understanding of experimentally observed nontrivial noise features and removes the need to perform hundreds of stochastic simulations to compute noise statistics. Our results pave the way towards understanding the sensitivity of noise to power-dependent physical processes such as soliton fission and Raman scattering, which in turn may be used to engineer pulse propagation to demonstrate controllable, on-demand quantum noise properties such as squeezing and entanglement.

%Nonlinear fibers have emerged as one of the key platforms for studying multimode nonlinear optics, providing a testbed to explore classical and quantum phenomena that emerge when multiple spatial and frequency modes interact in a nonlinear medium \cite{dudley2006supercontinuum}.  

\section{Dispersive Fourier transform experiment}

Our experimental setup is depicted in Fig. \ref{fig:1}a, where a femtosecond pulse experiences dispersion and Kerr nonlinearity in an optical fiber. The pump laser (centered around 1560 nm with 80 MHz repetition rate and pulse width $\sim 185$ fs) is operated near full power (through use of an amplifier that introduces $>30$ dB excess noise above the shot noise limit); experiments at different input powers are conducted using polarization-based attenuation. The femtosecond pulse propagates for $L=10$ m through a PM1550 fiber and its mean field spectrum is recorded using an optical spectrum analyzer (OSA). The pulse is then attenuated and passes through 4 km of a DFT fiber (SMF28). Chromatic dispersion stretches each pulse temporally and maps its spectrum onto a temporal waveform that is probed by a GHz-bandwidth oscilloscope at the output of the fiber. Details on the post-processing of the DFT data are provided in the Supplementary Information (SI).

\section{Generalized nonlinear Schrodinger equation and quantum sensitivity analysis}

To complement our experimental measurements, we numerically simulate the ultrafast pulse propagation using the generalized nonlinear Schrodinger equation (GNLSE), which in the mean field reads \cite{dudley2006supercontinuum, dudley2010supercontinuum, drummond2001quantum}
\begin{align}
\begin{split}
    \partial_z & A(z,t) = -\frac{\alpha}{2}A + i\sum_{k=2}^\infty \frac{i^k D_k}{k!}\frac{\partial^k}{\partial t^k}A(z,t) \\ &+ i\gamma\left(\int dt' R(t')|A(z,t-t')|^2\right)A(z,t),
\end{split}
\end{align}
where $A(z,t)$ denotes the field amplitude at position $z$ and retarded time $t$, $\alpha$ captures the loss in the fiber, $\beta_k$ are the dispersion coefficients, $\gamma$ represents the Kerr nonlinearity strength, and $R(t)=(1-f_R)\delta(t) + f_Rh_R(t)$ is the Raman response function ($f_R$ denotes the fractional Raman contribution). We use the well-known functional form $h_R(t)=\frac{\tau_1^2+\tau_2^2}{\tau_1\tau_2^2}\exp(-t/\tau_2)\sin(t/\tau_1)\theta(t)$, where $\theta(t)$ is the Heaviside step function (to ensure causality) and $\tau_1=0.012,\tau_2=0.032$ ps characterize the Raman gain spectrum for silica \cite{blow1989theoretical}.

To calculate noise properties, we augment the GNLSE with quantum sensitivity analysis (QSA), a recently proposed method to study the sensitivity of the noise in an output observable with respect to fluctuations in the input modes of a multimode system \cite{uddin2023ab}. The output intensity noise fluctuations can be calculated according to
\begin{align}
    \langle \delta n_\omega\delta n_{\omega'}\rangle = \int \frac{d\omega_1}{\pi} F_{\omega_1}\mathrm{Re}\left(\frac{\partial n_\omega}{\partial a_{\omega_1}}\frac{\partial n_{\omega'}}{\partial a^*_{\omega_1}}\right)
\end{align}
where $\delta n_\omega$ denotes the intensity fluctuation at frequency $\omega$, $a_{\omega_1}$ denotes the input pulse amplitude at frequency $\omega_1$, and $F_{\omega_1}$ denotes the excess noise at frequency $\omega_1$. Note that noise terms associated with Raman gain should also in theory be included through appropriate Langevin forces, but we have found the contribution from these terms to be negligible \cite{uddin2023ab}. We calculate $F_{\omega_1}$ by noting that the Fano factor of light after attenuation is given by $F_\mathrm{out}=(1-\alpha) F_\mathrm{in}+\alpha$ with $\alpha$ the absorption factor. $F_\mathrm{in}$ is given by the amplifier's noise spectrum, which we model by a Lorentzian centered around the pump wavelength at $\lambda\approx 1550$ nm. The calculation for $F_\mathrm{out}$ then directly gives the frequency-dependent excess input noise $F_{\omega_1}$. Finally, observe that the degenerate case $\omega=\omega'$ simply gives the intensity variance at frequency $\omega$. 

We briefly note that a common method to calculate noise statistics is based on simulating the GNLSE many times (often $N > 500$) with an appropriate noise profile sampled from a random distribution and added to the mean field input pulse for each simulation. We found that, while this method works for lower powers, it is not able to capture more intricate noise features at higher power. Also, in contrast to performing ensemble statistics (as in the first method), QSA calculates noise properties in a single pass by computing the Jacobian of the output intensity spectrum with respect to the input pulse using automatic differentiation.

Finally, in the results we show below, we model the noise floor of the DFT measurements by ``filtering out'' the noise for wavelengths with low intensity. This is motivated by the fact that statistics for a wavelength near the noise floor of the oscilloscope are not physically meaningful and will be governed by the noise of the instrument. Therefore, in computing single wavelength noise statistics, we set the noise for all wavelengths with intensities below a fixed minimum intensity to a fixed background value representing the noise floor. We also set to zero all noise correlations between wavelengths where at least one wavelength lies below the noise floor.

\begin{figure}
    \centering
    \includegraphics[width=\linewidth]{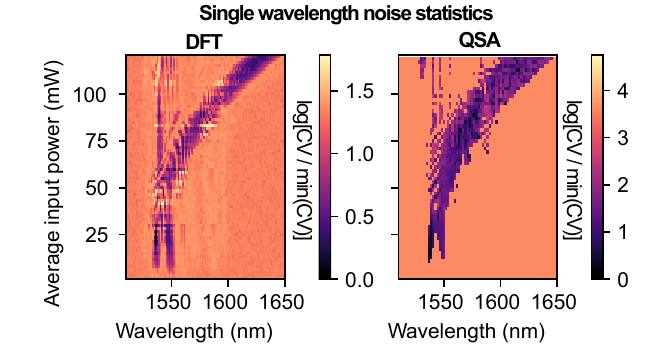}
    \caption{\textbf{Single wavelength noise statistics measured by DFT and calculated by QSA.} Coefficient of variation as a function of input power and wavelength, defined as $\mathrm{CV}\equiv\sqrt{\langle\delta n_\omega^2\rangle}/\langle n_\omega\rangle$ with $\sqrt{\langle\delta n_\omega^2\rangle}, \langle n_\omega\rangle$ the standard deviation and average power (respectively) at frequency $\omega$ ($\lambda=2\pi c/\omega$). Plotted is $\log{[\mathrm{CV}/\mathrm{min}(\mathrm{CV})]}$.}
    \label{fig:2}
\end{figure}

%(In practice, this noise profile was calculated by calculating the expected noise corresponding to a Fano factor $F_\mathrm{out}$ above the shot noise limit at each input wavelength.) We also added Langevin forces corresponding to the Raman gain processes but found that this source of noise was largely negligible compared to noise on the input pulse.

\section{Mean field behavior}

In Fig. \ref{fig:1}b, we compare the DFT mean field spectra extracted by averaging oscilloscope traces, the spectra measured by the OSA, and the spectra at the output of an $L=10$ m fiber simulated using the GNLSE, including effects of nonlinearity, dispersion, self-steepening, and Raman scattering. We observe ``breathing'' effects as well as the emergence of a Raman peak (Raman soliton) in the spectra at high power. The Raman soliton branches off at the fission point around 50 mW (average input power) and shifts to longer wavelengths as the power is increased. This shift is well-known and depends on the interplay between the strengths of nonlinearity and anomalous dispersion. Fig. \ref{fig:1}c shows the good agreement between the GNLSE simulation and the mean field measurement probed by OSA and DFT at high power, particularly for the position and shape of the Raman soliton (near 1625 nm at an input power of 120 mW).

The discrepancy between the mean field spectra calculated by OSA and DFT (e.g., the slight difference in wavelength of the Raman soliton peak) may be attributed to slight differences in the conversion between the temporal and frequency domains at different powers for the DFT data (see SI) or weak nonlinear effects in the DFT fiber.

\section{Single- and two-mode quantum statistics}

\begin{figure*}[t]
    \centering
    \includegraphics[width=\linewidth]{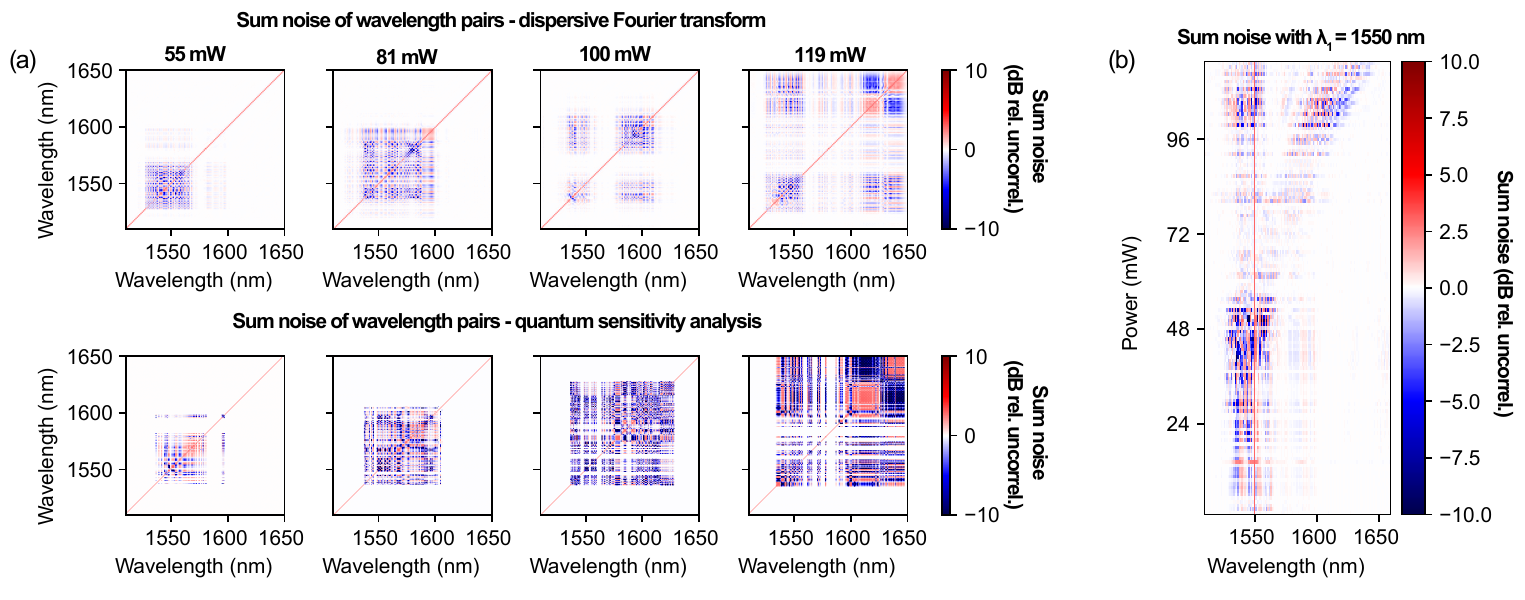}
    \caption{\textbf{Two-mode (twin wavelength) noise properties.} (a) Sum noise of wavelength pairs relative to uncorrelated sum noise $[ \langle \delta n_{\omega_1} + \delta n_{\omega_2}\rangle^2] / [\langle \delta n_{\omega_1}^2\rangle + \langle \delta n_{\omega_2}^2\rangle]$, where $\lambda_{1,2}=2\pi c /\omega_{1,2}$, for experimental data using dispersive Fourier transform (top row) and theoretical quantum sensitivity analysis (bottom row). (b) Sum noise with one wavelength in the pair fixed at $\lambda_1=1550$ nm as a function of power and the second wavelength $\lambda_2$.}
    \label{fig:3}
\end{figure*}

We now turn our attention to the noise properties in this system. We begin by examining noise at individual wavelengths. In Fig. \ref{fig:2}, we plot the coefficient of variation (CV), the ratio of standard deviation to mean intensity, as a function of input power and (output) wavelength. Of note is the low noise of the Raman branch, which suggests that Raman scattering does not significantly amplify noise even at high power (where our setup's amplifier adds a large amount of excess noise around the input wavelength at 1550 nm) even though the Raman branch has a large average intensity (see Fig. \ref{fig:1}b/c). 
%The soliton fission point around 50 mW is accompanied by high noise around the input wavelength, however.

We next examine noise correlations between pairs of wavelengths at different input powers. In Fig. \ref{fig:3}a, we plot the sum noise relative to the case of uncorrelated wavelength pairs, $[ \langle \delta n_{\omega_1} + \delta n_{\omega_2}\rangle^2] / [\langle \delta n_{\omega_1}^2\rangle + \langle \delta n_{\omega_2}^2\rangle]$ where $\lambda_{1,2}=2\pi c /\omega_{1,2}$. Regions in red indicate antisqueezing above the uncorrelated limit, while regions in blue indicate squeezing below the uncorrelated limit and imply entangled wavelength pairs. At low powers, no nontrivial correlations are present. For input powers around 50 mW, strong correlations begin to develop around the pump wavelength, associated with four-wave mixing processes in the generation of the fundamental soliton. The checkerboard correlation map observed has been attributed to self-phase modulation and wavelength jitter \cite{godin2013real}. As the power increases, strong correlations shift to longer wavelengths in accordance with Raman scattering. The correlation structure around the pump wavelength remains, but a similar structure also emerges around the Raman peak. Furthermore, longer-range entanglement between pump and Raman wavelengths emerges. Again, these correlations remain robust despite the large amount of excess noise introduced by the setup's amplifier. At high power, we observe high sensitivity to small changes in input power for single- and two-mode noise properties (see SI), which may be attributed to interference effects from the power-sensitive nonlinear phase shift.

QSA provides reasonable agreement with the experimentally observed regions of strong correlation. QSA predicts finer features in the correlation heatmaps, but we suspect that, due to the noise floor and limited resolution of the oscilloscope, these features are ``washed out'' in the experimental data (which we model in our simulations by imposing a noise floor below which correlations are neglected).

In Fig. \ref{fig:3}b, we plot the sum noise with one wavelength fixed at $\lambda=1550$ nm as a function of input power. At low power, weak entanglement exists with wavelengths within the pump region. After the soliton fission point in which the Raman soliton breaks off, entanglement features emerge with the pump and Raman solitons. The correlations around the pump wavelength are strongest near the fission point, are weaker in the region between 60 and 100 mW past the fission point, and re-emerge past 100 mW. 

\begin{figure*}[t]
    \centering
    \includegraphics[width=\linewidth]{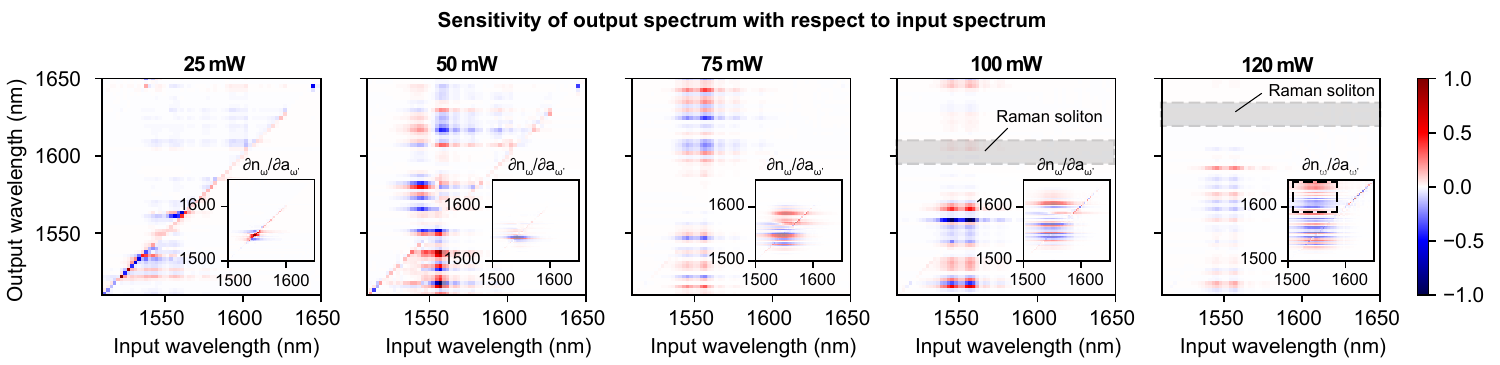}
    \caption{\textbf{Sensitivity of output spectrum with respect to input spectrum.} Plots of the normalized Jacobian $J_{\omega,\omega_1}=\frac{a_{\omega_1}}{n_\omega}\frac{\partial n_\omega}{\partial a_{\omega_1}}$ at five different input powers, where $n_\omega$ denotes the output intensity at frequency $\omega$ and $a_{\omega_1}$ denotes the input pulse amplitude at frequency $\omega_1$, using automatic differentiation on the GNLSE. To highlight the sign flip in the Jacobian across the Raman soliton at high power, $\frac{\partial n_\omega}{\partial a_{\omega_1}}$ is plotted in the insets. The normalization in all plots is to the maximum value at the corresponding input power.}
    \label{fig:4}
\end{figure*}

Finally, in Fig. \ref{fig:4}, we explore the sensitivity of the output spectrum to fluctuations in the input pulse. We plot a normalized version of the Jacobian $J_{\omega,\omega_1}=\frac{a_{\omega_1}}{n_\omega}\frac{\partial n_\omega}{\partial a_{\omega_1}}$, where $n_\omega$ denotes the output intensity at frequency $\omega$ and $a_{\omega_1}$ denotes the input pulse amplitude at frequency $\omega_1$. When examining noise properties at individual wavelengths, larger $|J_{\omega,\omega_1}|$ corresponds to greater sensitivity of the output intensity at $\omega$ to fluctuations in the input intensity at $\omega_1$. Therefore, larger $|J_{\omega,\omega_1}|$ at $\omega_1$ with high input noise (e.g., around the pump frequency) will give rise to larger output intensity noise at $\omega$. For example, we see that the Raman soliton has low $|J_{\omega,\omega_1}|$ for $2\pi c/\omega_1\approx 1550$ nm (the pump wavelength), where the amplifier introduces large amounts of excess noise, and therefore frequencies $\omega$ corresponding to the Raman soliton can show low noise. 

For wavelength pairs, the sensitivity analysis indicates where strong squeezing/anti-squeezing may be expected. For example, at the highest input powers, $J_{\omega,\omega_1}$ switches sign across the Raman soliton peak for $\omega_1$ near the pump frequency (see dashed box in inset of Fig. \ref{fig:4}). Therefore, constructive interference occurs for wavelength pairs in the same region, $\frac{\partial n_\omega}{\partial a_{\omega_1}}\frac{\partial n_{\omega'}}{\partial a_{\omega_1}}>0$, while destructive interference occurs for wavelengths chosen in different regions, $\frac{\partial n_\omega}{\partial a_{\omega_1}}\frac{\partial n_{\omega'}}{\partial a_{\omega_1}}<0$. This results in the observation of high noise along the diagonal and low noise along the anti-diagonal in Fig. \ref{fig:3}a at highest power in the Raman soliton region. Physically, these behaviors likely emerge from wavelength jitter in the Raman soliton \cite{godin2013real}.

\section{Discussion and outlook}

In this paper, we have used the dispersive Fourier transform (DFT) technique to probe the power-dependent mean field and noise properties of ultrafast pulses propagating through a nonlinear fiber. DFT shows good agreement with the mean field spectrum measured by a conventional optical spectrum analyzer and the spectrum predicted by a generalized nonlinear Schrodinger equation (GNLSE) analysis. Single wavelength noise statistics on the spectra collected by DFT suggest a low noise Raman soliton that progressively redshifts as the input power is increased. Twin wavelength correlations indicate the emergence of entanglement within the Raman soliton and between the Raman and pump wavelengths, with the region in between these two wavelengths having low intensity. These results are supported by a quantum sensitivity analysis (QSA), which reveals that low noise features emerge because of a weak sensitivity to noise on the input pulse at particular wavelengths, making these features insensitive to the large amount of excess noise introduced by the setup's amplifier. QSA predicts finer noise features (with potentially strong entanglement features), particularly in the depletion region, that may be picked up using an oscilloscope with finer amplitude resolution and lower noise floor.

Compared to traditional noise measurements using an electronic spectrum analyzer (ESA), DFT offers finer wavelength resolution ($\sim 0.1$ nm versus $\sim 1$ nm \cite{uddin2023ab}), allows direct calculation of higher-order statistical moments, and enables automatic calculation of noise correlations between all possible wavelength pairs in a single experiment (in ESA, by contrast, correlations between different wavelength pairs must be measured sequentially). In combination with quantum sensitivity analysis (QSA), DFT can permit a clearer understanding of the physical processes responsible for various noise behaviors and how the input pulse can be modified to achieve desired quantum noise properties. Our results for soliton fission and Raman scattering here suggest that DFT and QSA are generalizable techniques for understanding the classical and quantum properties of several phenomena in multimode nonlinear optics, including ultrafast modal dispersion in multimode fibers. We anticipate that our results will establish the combination of DFT and QSA as a valuable tool for probing multimode quantum noise and the emergence of novel quantum states generated by the nonlinear interactions of ultrafast pulses.

\section{Acknowledgements} S.P. acknowledges the financial support of the Hertz Fellowship Program and NSF Graduate Research Fellowship Program. This material is based upon work supported in part by the Air Force Office of Scientific Research under the award number FA9550-20-1-0115 and DARPA Agreement No. HO0011249049; the work is also supported in part by the U. S. Army Research Office through the Institute for Soldier Nanotechnologies at MIT, under Collaborative Agreement Number W911NF-23-2-0121. We also acknowledge support of Parviz Tayebati.

\bibliography{main}

\begin{thebibliography}{38}%
\makeatletter
\providecommand \@ifxundefined [1]{%
 \@ifx{#1\undefined}
}%
\providecommand \@ifnum [1]{%
 \ifnum #1\expandafter \@firstoftwo
 \else \expandafter \@secondoftwo
 \fi
}%
\providecommand \@ifx [1]{%
 \ifx #1\expandafter \@firstoftwo
 \else \expandafter \@secondoftwo
 \fi
}%
\providecommand \natexlab [1]{#1}%
\providecommand \enquote  [1]{``#1''}%
\providecommand \bibnamefont  [1]{#1}%
\providecommand \bibfnamefont [1]{#1}%
\providecommand \citenamefont [1]{#1}%
\providecommand \href@noop [0]{\@secondoftwo}%
\providecommand \href [0]{\begingroup \@sanitize@url \@href}%
\providecommand \@href[1]{\@@startlink{#1}\@@href}%
\providecommand \@@href[1]{\endgroup#1\@@endlink}%
\providecommand \@sanitize@url [0]{\catcode `\\12\catcode `\$12\catcode `\&12\catcode `\#12\catcode `\^12\catcode `\_12\catcode `\%12\relax}%
\providecommand \@@startlink[1]{}%
\providecommand \@@endlink[0]{}%
\providecommand \url  [0]{\begingroup\@sanitize@url \@url }%
\providecommand \@url [1]{\endgroup\@href {#1}{\urlprefix }}%
\providecommand \urlprefix  [0]{URL }%
\providecommand \Eprint [0]{\href }%
\providecommand \doibase [0]{https://doi.org/}%
\providecommand \selectlanguage [0]{\@gobble}%
\providecommand \bibinfo  [0]{\@secondoftwo}%
\providecommand \bibfield  [0]{\@secondoftwo}%
\providecommand \translation [1]{[#1]}%
\providecommand \BibitemOpen [0]{}%
\providecommand \bibitemStop [0]{}%
\providecommand \bibitemNoStop [0]{.\EOS\space}%
\providecommand \EOS [0]{\spacefactor3000\relax}%
\providecommand \BibitemShut  [1]{\csname bibitem#1\endcsname}%
\let\auto@bib@innerbib\@empty
%</preamble>
\bibitem [{\citenamefont {Dang}\ \emph {et~al.}(2024)\citenamefont {Dang}, \citenamefont {Deng}, \citenamefont {Chen}, \citenamefont {Ding},\ and\ \citenamefont {Zhang}}]{dang2024simple}%
  \BibitemOpen
  \bibfield  {author} {\bibinfo {author} {\bibfnamefont {Z.}~\bibnamefont {Dang}}, \bibinfo {author} {\bibfnamefont {Z.}~\bibnamefont {Deng}}, \bibinfo {author} {\bibfnamefont {T.}~\bibnamefont {Chen}}, \bibinfo {author} {\bibfnamefont {Z.}~\bibnamefont {Ding}},\ and\ \bibinfo {author} {\bibfnamefont {Z.}~\bibnamefont {Zhang}},\ }\bibfield  {title} {\bibinfo {title} {A simple nonlinear classifier using a multimode optical chip},\ }\href@noop {} {\bibfield  {journal} {\bibinfo  {journal} {Advanced Photonics Research}\ }\textbf {\bibinfo {volume} {5}},\ \bibinfo {pages} {2300253} (\bibinfo {year} {2024})}\BibitemShut {NoStop}%
\bibitem [{\citenamefont {Oguz}\ \emph {et~al.}(2024)\citenamefont {Oguz}, \citenamefont {Hsieh}, \citenamefont {Dinc}, \citenamefont {Te{\u{g}}in}, \citenamefont {Yildirim}, \citenamefont {Gigli}, \citenamefont {Moser},\ and\ \citenamefont {Psaltis}}]{oguz2024programming}%
  \BibitemOpen
  \bibfield  {author} {\bibinfo {author} {\bibfnamefont {I.}~\bibnamefont {Oguz}}, \bibinfo {author} {\bibfnamefont {J.-L.}\ \bibnamefont {Hsieh}}, \bibinfo {author} {\bibfnamefont {N.~U.}\ \bibnamefont {Dinc}}, \bibinfo {author} {\bibfnamefont {U.}~\bibnamefont {Te{\u{g}}in}}, \bibinfo {author} {\bibfnamefont {M.}~\bibnamefont {Yildirim}}, \bibinfo {author} {\bibfnamefont {C.}~\bibnamefont {Gigli}}, \bibinfo {author} {\bibfnamefont {C.}~\bibnamefont {Moser}},\ and\ \bibinfo {author} {\bibfnamefont {D.}~\bibnamefont {Psaltis}},\ }\bibfield  {title} {\bibinfo {title} {Programming nonlinear propagation for efficient optical learning machines},\ }\href@noop {} {\bibfield  {journal} {\bibinfo  {journal} {Advanced Photonics}\ }\textbf {\bibinfo {volume} {6}},\ \bibinfo {pages} {016002} (\bibinfo {year} {2024})}\BibitemShut {NoStop}%
\bibitem [{\citenamefont {Yildirim}\ \emph {et~al.}(2023)\citenamefont {Yildirim}, \citenamefont {Oguz}, \citenamefont {Kaufmann}, \citenamefont {Escal{\'e}}, \citenamefont {Grange}, \citenamefont {Psaltis},\ and\ \citenamefont {Moser}}]{yildirim2023nonlinear}%
  \BibitemOpen
  \bibfield  {author} {\bibinfo {author} {\bibfnamefont {M.}~\bibnamefont {Yildirim}}, \bibinfo {author} {\bibfnamefont {I.}~\bibnamefont {Oguz}}, \bibinfo {author} {\bibfnamefont {F.}~\bibnamefont {Kaufmann}}, \bibinfo {author} {\bibfnamefont {M.~R.}\ \bibnamefont {Escal{\'e}}}, \bibinfo {author} {\bibfnamefont {R.}~\bibnamefont {Grange}}, \bibinfo {author} {\bibfnamefont {D.}~\bibnamefont {Psaltis}},\ and\ \bibinfo {author} {\bibfnamefont {C.}~\bibnamefont {Moser}},\ }\bibfield  {title} {\bibinfo {title} {Nonlinear optical feature generator for machine learning},\ }\href@noop {} {\bibfield  {journal} {\bibinfo  {journal} {Apl Photonics}\ }\textbf {\bibinfo {volume} {8}} (\bibinfo {year} {2023})}\BibitemShut {NoStop}%
\bibitem [{\citenamefont {Rahmani}\ \emph {et~al.}(2022)\citenamefont {Rahmani}, \citenamefont {Oguz}, \citenamefont {Tegin}, \citenamefont {Hsieh}, \citenamefont {Psaltis},\ and\ \citenamefont {Moser}}]{rahmani2022learning}%
  \BibitemOpen
  \bibfield  {author} {\bibinfo {author} {\bibfnamefont {B.}~\bibnamefont {Rahmani}}, \bibinfo {author} {\bibfnamefont {I.}~\bibnamefont {Oguz}}, \bibinfo {author} {\bibfnamefont {U.}~\bibnamefont {Tegin}}, \bibinfo {author} {\bibfnamefont {J.-l.}\ \bibnamefont {Hsieh}}, \bibinfo {author} {\bibfnamefont {D.}~\bibnamefont {Psaltis}},\ and\ \bibinfo {author} {\bibfnamefont {C.}~\bibnamefont {Moser}},\ }\bibfield  {title} {\bibinfo {title} {Learning to image and compute with multimode optical fibers},\ }\href@noop {} {\bibfield  {journal} {\bibinfo  {journal} {Nanophotonics}\ }\textbf {\bibinfo {volume} {11}},\ \bibinfo {pages} {1071} (\bibinfo {year} {2022})}\BibitemShut {NoStop}%
\bibitem [{\citenamefont {Sunada}\ \emph {et~al.}(2020)\citenamefont {Sunada}, \citenamefont {Kanno},\ and\ \citenamefont {Uchida}}]{sunada2020using}%
  \BibitemOpen
  \bibfield  {author} {\bibinfo {author} {\bibfnamefont {S.}~\bibnamefont {Sunada}}, \bibinfo {author} {\bibfnamefont {K.}~\bibnamefont {Kanno}},\ and\ \bibinfo {author} {\bibfnamefont {A.}~\bibnamefont {Uchida}},\ }\bibfield  {title} {\bibinfo {title} {Using multidimensional speckle dynamics for high-speed, large-scale, parallel photonic computing},\ }\href@noop {} {\bibfield  {journal} {\bibinfo  {journal} {Optics Express}\ }\textbf {\bibinfo {volume} {28}},\ \bibinfo {pages} {30349} (\bibinfo {year} {2020})}\BibitemShut {NoStop}%
\bibitem [{\citenamefont {Cristiani}\ \emph {et~al.}(2022)\citenamefont {Cristiani}, \citenamefont {Lacava}, \citenamefont {Rademacher}, \citenamefont {Puttnam}, \citenamefont {Lu{\`\i}s}, \citenamefont {Antonelli}, \citenamefont {Mecozzi}, \citenamefont {Shtaif}, \citenamefont {Cozzolino}, \citenamefont {Bacco} \emph {et~al.}}]{cristiani2022roadmap}%
  \BibitemOpen
  \bibfield  {author} {\bibinfo {author} {\bibfnamefont {I.}~\bibnamefont {Cristiani}}, \bibinfo {author} {\bibfnamefont {C.}~\bibnamefont {Lacava}}, \bibinfo {author} {\bibfnamefont {G.}~\bibnamefont {Rademacher}}, \bibinfo {author} {\bibfnamefont {B.~J.}\ \bibnamefont {Puttnam}}, \bibinfo {author} {\bibfnamefont {R.~S.}\ \bibnamefont {Lu{\`\i}s}}, \bibinfo {author} {\bibfnamefont {C.}~\bibnamefont {Antonelli}}, \bibinfo {author} {\bibfnamefont {A.}~\bibnamefont {Mecozzi}}, \bibinfo {author} {\bibfnamefont {M.}~\bibnamefont {Shtaif}}, \bibinfo {author} {\bibfnamefont {D.}~\bibnamefont {Cozzolino}}, \bibinfo {author} {\bibfnamefont {D.}~\bibnamefont {Bacco}}, \emph {et~al.},\ }\bibfield  {title} {\bibinfo {title} {Roadmap on multimode photonics},\ }\href@noop {} {\bibfield  {journal} {\bibinfo  {journal} {Journal of Optics}\ }\textbf {\bibinfo {volume} {24}},\ \bibinfo {pages} {083001} (\bibinfo {year} {2022})}\BibitemShut {NoStop}%
\bibitem [{\citenamefont {Wright}\ \emph {et~al.}(2022{\natexlab{a}})\citenamefont {Wright}, \citenamefont {Renninger}, \citenamefont {Christodoulides},\ and\ \citenamefont {Wise}}]{wright2022nonlinear}%
  \BibitemOpen
  \bibfield  {author} {\bibinfo {author} {\bibfnamefont {L.~G.}\ \bibnamefont {Wright}}, \bibinfo {author} {\bibfnamefont {W.~H.}\ \bibnamefont {Renninger}}, \bibinfo {author} {\bibfnamefont {D.~N.}\ \bibnamefont {Christodoulides}},\ and\ \bibinfo {author} {\bibfnamefont {F.~W.}\ \bibnamefont {Wise}},\ }\bibfield  {title} {\bibinfo {title} {Nonlinear multimode photonics: nonlinear optics with many degrees of freedom},\ }\href@noop {} {\bibfield  {journal} {\bibinfo  {journal} {Optica}\ }\textbf {\bibinfo {volume} {9}},\ \bibinfo {pages} {824} (\bibinfo {year} {2022}{\natexlab{a}})}\BibitemShut {NoStop}%
\bibitem [{\citenamefont {Yanagimoto}\ \emph {et~al.}(2023)\citenamefont {Yanagimoto}, \citenamefont {Ng}, \citenamefont {Jankowski}, \citenamefont {Nehra}, \citenamefont {McKenna}, \citenamefont {Onodera}, \citenamefont {Wright}, \citenamefont {Hamerly}, \citenamefont {Marandi}, \citenamefont {Fejer} \emph {et~al.}}]{yanagimoto2023mesoscopic}%
  \BibitemOpen
  \bibfield  {author} {\bibinfo {author} {\bibfnamefont {R.}~\bibnamefont {Yanagimoto}}, \bibinfo {author} {\bibfnamefont {E.}~\bibnamefont {Ng}}, \bibinfo {author} {\bibfnamefont {M.}~\bibnamefont {Jankowski}}, \bibinfo {author} {\bibfnamefont {R.}~\bibnamefont {Nehra}}, \bibinfo {author} {\bibfnamefont {T.~P.}\ \bibnamefont {McKenna}}, \bibinfo {author} {\bibfnamefont {T.}~\bibnamefont {Onodera}}, \bibinfo {author} {\bibfnamefont {L.~G.}\ \bibnamefont {Wright}}, \bibinfo {author} {\bibfnamefont {R.}~\bibnamefont {Hamerly}}, \bibinfo {author} {\bibfnamefont {A.}~\bibnamefont {Marandi}}, \bibinfo {author} {\bibfnamefont {M.}~\bibnamefont {Fejer}}, \emph {et~al.},\ }\bibfield  {title} {\bibinfo {title} {Mesoscopic ultrafast nonlinear optics--the emergence of multimode quantum non-gaussian physics},\ }\href@noop {} {\bibfield  {journal} {\bibinfo  {journal} {arXiv preprint arXiv:2311.13775}\ } (\bibinfo {year} {2023})}\BibitemShut {NoStop}%
\bibitem [{\citenamefont {Guidry}\ \emph {et~al.}(2022)\citenamefont {Guidry}, \citenamefont {Lukin}, \citenamefont {Yang}, \citenamefont {Trivedi},\ and\ \citenamefont {Vu{\v{c}}kovi{\'c}}}]{guidry2022quantum}%
  \BibitemOpen
  \bibfield  {author} {\bibinfo {author} {\bibfnamefont {M.~A.}\ \bibnamefont {Guidry}}, \bibinfo {author} {\bibfnamefont {D.~M.}\ \bibnamefont {Lukin}}, \bibinfo {author} {\bibfnamefont {K.~Y.}\ \bibnamefont {Yang}}, \bibinfo {author} {\bibfnamefont {R.}~\bibnamefont {Trivedi}},\ and\ \bibinfo {author} {\bibfnamefont {J.}~\bibnamefont {Vu{\v{c}}kovi{\'c}}},\ }\bibfield  {title} {\bibinfo {title} {Quantum optics of soliton microcombs},\ }\href@noop {} {\bibfield  {journal} {\bibinfo  {journal} {Nature Photonics}\ }\textbf {\bibinfo {volume} {16}},\ \bibinfo {pages} {52} (\bibinfo {year} {2022})}\BibitemShut {NoStop}%
\bibitem [{\citenamefont {Guidry}\ \emph {et~al.}(2023)\citenamefont {Guidry}, \citenamefont {Lukin}, \citenamefont {Yang},\ and\ \citenamefont {Vu{\v{c}}kovi{\'c}}}]{guidry2023multimode}%
  \BibitemOpen
  \bibfield  {author} {\bibinfo {author} {\bibfnamefont {M.~A.}\ \bibnamefont {Guidry}}, \bibinfo {author} {\bibfnamefont {D.~M.}\ \bibnamefont {Lukin}}, \bibinfo {author} {\bibfnamefont {K.~Y.}\ \bibnamefont {Yang}},\ and\ \bibinfo {author} {\bibfnamefont {J.}~\bibnamefont {Vu{\v{c}}kovi{\'c}}},\ }\bibfield  {title} {\bibinfo {title} {Multimode squeezing in soliton crystal microcombs},\ }\href@noop {} {\bibfield  {journal} {\bibinfo  {journal} {Optica}\ }\textbf {\bibinfo {volume} {10}},\ \bibinfo {pages} {694} (\bibinfo {year} {2023})}\BibitemShut {NoStop}%
\bibitem [{\citenamefont {Lustig}\ \emph {et~al.}(2024)\citenamefont {Lustig}, \citenamefont {Guidry}, \citenamefont {Lukin}, \citenamefont {Fan},\ and\ \citenamefont {Vuckovic}}]{lustig2024emerging}%
  \BibitemOpen
  \bibfield  {author} {\bibinfo {author} {\bibfnamefont {E.}~\bibnamefont {Lustig}}, \bibinfo {author} {\bibfnamefont {M.~A.}\ \bibnamefont {Guidry}}, \bibinfo {author} {\bibfnamefont {D.~M.}\ \bibnamefont {Lukin}}, \bibinfo {author} {\bibfnamefont {S.}~\bibnamefont {Fan}},\ and\ \bibinfo {author} {\bibfnamefont {J.}~\bibnamefont {Vuckovic}},\ }\bibfield  {title} {\bibinfo {title} {Emerging quadrature lattices of kerr combs},\ }\href@noop {} {\bibfield  {journal} {\bibinfo  {journal} {arXiv preprint arXiv:2407.13049}\ } (\bibinfo {year} {2024})}\BibitemShut {NoStop}%
\bibitem [{\citenamefont {Kues}\ \emph {et~al.}(2019)\citenamefont {Kues}, \citenamefont {Reimer}, \citenamefont {Lukens}, \citenamefont {Munro}, \citenamefont {Weiner}, \citenamefont {Moss},\ and\ \citenamefont {Morandotti}}]{kues2019quantum}%
  \BibitemOpen
  \bibfield  {author} {\bibinfo {author} {\bibfnamefont {M.}~\bibnamefont {Kues}}, \bibinfo {author} {\bibfnamefont {C.}~\bibnamefont {Reimer}}, \bibinfo {author} {\bibfnamefont {J.~M.}\ \bibnamefont {Lukens}}, \bibinfo {author} {\bibfnamefont {W.~J.}\ \bibnamefont {Munro}}, \bibinfo {author} {\bibfnamefont {A.~M.}\ \bibnamefont {Weiner}}, \bibinfo {author} {\bibfnamefont {D.~J.}\ \bibnamefont {Moss}},\ and\ \bibinfo {author} {\bibfnamefont {R.}~\bibnamefont {Morandotti}},\ }\bibfield  {title} {\bibinfo {title} {Quantum optical microcombs},\ }\href@noop {} {\bibfield  {journal} {\bibinfo  {journal} {Nature Photonics}\ }\textbf {\bibinfo {volume} {13}},\ \bibinfo {pages} {170} (\bibinfo {year} {2019})}\BibitemShut {NoStop}%
\bibitem [{\citenamefont {Roslund}\ \emph {et~al.}(2014)\citenamefont {Roslund}, \citenamefont {De~Araujo}, \citenamefont {Jiang}, \citenamefont {Fabre},\ and\ \citenamefont {Treps}}]{roslund2014wavelength}%
  \BibitemOpen
  \bibfield  {author} {\bibinfo {author} {\bibfnamefont {J.}~\bibnamefont {Roslund}}, \bibinfo {author} {\bibfnamefont {R.~M.}\ \bibnamefont {De~Araujo}}, \bibinfo {author} {\bibfnamefont {S.}~\bibnamefont {Jiang}}, \bibinfo {author} {\bibfnamefont {C.}~\bibnamefont {Fabre}},\ and\ \bibinfo {author} {\bibfnamefont {N.}~\bibnamefont {Treps}},\ }\bibfield  {title} {\bibinfo {title} {Wavelength-multiplexed quantum networks with ultrafast frequency combs},\ }\href@noop {} {\bibfield  {journal} {\bibinfo  {journal} {Nature Photonics}\ }\textbf {\bibinfo {volume} {8}},\ \bibinfo {pages} {109} (\bibinfo {year} {2014})}\BibitemShut {NoStop}%
\bibitem [{\citenamefont {Ra}\ \emph {et~al.}(2020)\citenamefont {Ra}, \citenamefont {Dufour}, \citenamefont {Walschaers}, \citenamefont {Jacquard}, \citenamefont {Michel}, \citenamefont {Fabre},\ and\ \citenamefont {Treps}}]{ra2020non}%
  \BibitemOpen
  \bibfield  {author} {\bibinfo {author} {\bibfnamefont {Y.-S.}\ \bibnamefont {Ra}}, \bibinfo {author} {\bibfnamefont {A.}~\bibnamefont {Dufour}}, \bibinfo {author} {\bibfnamefont {M.}~\bibnamefont {Walschaers}}, \bibinfo {author} {\bibfnamefont {C.}~\bibnamefont {Jacquard}}, \bibinfo {author} {\bibfnamefont {T.}~\bibnamefont {Michel}}, \bibinfo {author} {\bibfnamefont {C.}~\bibnamefont {Fabre}},\ and\ \bibinfo {author} {\bibfnamefont {N.}~\bibnamefont {Treps}},\ }\bibfield  {title} {\bibinfo {title} {Non-gaussian quantum states of a multimode light field},\ }\href@noop {} {\bibfield  {journal} {\bibinfo  {journal} {Nature Physics}\ }\textbf {\bibinfo {volume} {16}},\ \bibinfo {pages} {144} (\bibinfo {year} {2020})}\BibitemShut {NoStop}%
\bibitem [{\citenamefont {Presutti}\ \emph {et~al.}(2024)\citenamefont {Presutti}, \citenamefont {Wright}, \citenamefont {Ma}, \citenamefont {Wang}, \citenamefont {Malia}, \citenamefont {Onodera},\ and\ \citenamefont {McMahon}}]{presutti2024highly}%
  \BibitemOpen
  \bibfield  {author} {\bibinfo {author} {\bibfnamefont {F.}~\bibnamefont {Presutti}}, \bibinfo {author} {\bibfnamefont {L.~G.}\ \bibnamefont {Wright}}, \bibinfo {author} {\bibfnamefont {S.-Y.}\ \bibnamefont {Ma}}, \bibinfo {author} {\bibfnamefont {T.}~\bibnamefont {Wang}}, \bibinfo {author} {\bibfnamefont {B.~K.}\ \bibnamefont {Malia}}, \bibinfo {author} {\bibfnamefont {T.}~\bibnamefont {Onodera}},\ and\ \bibinfo {author} {\bibfnamefont {P.~L.}\ \bibnamefont {McMahon}},\ }\bibfield  {title} {\bibinfo {title} {Highly multimode visible squeezed light with programmable spectral correlations through broadband up-conversion},\ }\href@noop {} {\bibfield  {journal} {\bibinfo  {journal} {arXiv preprint arXiv:2401.06119}\ } (\bibinfo {year} {2024})}\BibitemShut {NoStop}%
\bibitem [{\citenamefont {Yang}\ \emph {et~al.}(2021)\citenamefont {Yang}, \citenamefont {Jahanbozorgi}, \citenamefont {Jeong}, \citenamefont {Sun}, \citenamefont {Pfister}, \citenamefont {Lee},\ and\ \citenamefont {Yi}}]{yang2021squeezed}%
  \BibitemOpen
  \bibfield  {author} {\bibinfo {author} {\bibfnamefont {Z.}~\bibnamefont {Yang}}, \bibinfo {author} {\bibfnamefont {M.}~\bibnamefont {Jahanbozorgi}}, \bibinfo {author} {\bibfnamefont {D.}~\bibnamefont {Jeong}}, \bibinfo {author} {\bibfnamefont {S.}~\bibnamefont {Sun}}, \bibinfo {author} {\bibfnamefont {O.}~\bibnamefont {Pfister}}, \bibinfo {author} {\bibfnamefont {H.}~\bibnamefont {Lee}},\ and\ \bibinfo {author} {\bibfnamefont {X.}~\bibnamefont {Yi}},\ }\bibfield  {title} {\bibinfo {title} {A squeezed quantum microcomb on a chip},\ }\href@noop {} {\bibfield  {journal} {\bibinfo  {journal} {Nature Communications}\ }\textbf {\bibinfo {volume} {12}},\ \bibinfo {pages} {4781} (\bibinfo {year} {2021})}\BibitemShut {NoStop}%
\bibitem [{\citenamefont {Wright}\ \emph {et~al.}(2022{\natexlab{b}})\citenamefont {Wright}, \citenamefont {Wu}, \citenamefont {Christodoulides},\ and\ \citenamefont {Wise}}]{wright2022physics}%
  \BibitemOpen
  \bibfield  {author} {\bibinfo {author} {\bibfnamefont {L.~G.}\ \bibnamefont {Wright}}, \bibinfo {author} {\bibfnamefont {F.~O.}\ \bibnamefont {Wu}}, \bibinfo {author} {\bibfnamefont {D.~N.}\ \bibnamefont {Christodoulides}},\ and\ \bibinfo {author} {\bibfnamefont {F.~W.}\ \bibnamefont {Wise}},\ }\bibfield  {title} {\bibinfo {title} {Physics of highly multimode nonlinear optical systems},\ }\href@noop {} {\bibfield  {journal} {\bibinfo  {journal} {Nature Physics}\ }\textbf {\bibinfo {volume} {18}},\ \bibinfo {pages} {1018} (\bibinfo {year} {2022}{\natexlab{b}})}\BibitemShut {NoStop}%
\bibitem [{\citenamefont {Wu}\ \emph {et~al.}(2019)\citenamefont {Wu}, \citenamefont {Hassan},\ and\ \citenamefont {Christodoulides}}]{wu2019thermodynamic}%
  \BibitemOpen
  \bibfield  {author} {\bibinfo {author} {\bibfnamefont {F.~O.}\ \bibnamefont {Wu}}, \bibinfo {author} {\bibfnamefont {A.~U.}\ \bibnamefont {Hassan}},\ and\ \bibinfo {author} {\bibfnamefont {D.~N.}\ \bibnamefont {Christodoulides}},\ }\bibfield  {title} {\bibinfo {title} {Thermodynamic theory of highly multimoded nonlinear optical systems},\ }\href@noop {} {\bibfield  {journal} {\bibinfo  {journal} {Nature Photonics}\ }\textbf {\bibinfo {volume} {13}},\ \bibinfo {pages} {776} (\bibinfo {year} {2019})}\BibitemShut {NoStop}%
\bibitem [{\citenamefont {Makris}\ \emph {et~al.}(2020)\citenamefont {Makris}, \citenamefont {Wu}, \citenamefont {Jung},\ and\ \citenamefont {Christodoulides}}]{makris2020statistical}%
  \BibitemOpen
  \bibfield  {author} {\bibinfo {author} {\bibfnamefont {K.~G.}\ \bibnamefont {Makris}}, \bibinfo {author} {\bibfnamefont {F.~O.}\ \bibnamefont {Wu}}, \bibinfo {author} {\bibfnamefont {P.~S.}\ \bibnamefont {Jung}},\ and\ \bibinfo {author} {\bibfnamefont {D.~N.}\ \bibnamefont {Christodoulides}},\ }\bibfield  {title} {\bibinfo {title} {Statistical mechanics of weakly nonlinear optical multimode gases},\ }\href@noop {} {\bibfield  {journal} {\bibinfo  {journal} {Optics letters}\ }\textbf {\bibinfo {volume} {45}},\ \bibinfo {pages} {1651} (\bibinfo {year} {2020})}\BibitemShut {NoStop}%
\bibitem [{\citenamefont {Ferraro}\ \emph {et~al.}(2024)\citenamefont {Ferraro}, \citenamefont {Mangini}, \citenamefont {Wu}, \citenamefont {Zitelli}, \citenamefont {Christodoulides},\ and\ \citenamefont {Wabnitz}}]{ferraro2024calorimetry}%
  \BibitemOpen
  \bibfield  {author} {\bibinfo {author} {\bibfnamefont {M.}~\bibnamefont {Ferraro}}, \bibinfo {author} {\bibfnamefont {F.}~\bibnamefont {Mangini}}, \bibinfo {author} {\bibfnamefont {F.}~\bibnamefont {Wu}}, \bibinfo {author} {\bibfnamefont {M.}~\bibnamefont {Zitelli}}, \bibinfo {author} {\bibfnamefont {D.}~\bibnamefont {Christodoulides}},\ and\ \bibinfo {author} {\bibfnamefont {S.}~\bibnamefont {Wabnitz}},\ }\bibfield  {title} {\bibinfo {title} {Calorimetry of photon gases in nonlinear multimode optical fibers},\ }\href@noop {} {\bibfield  {journal} {\bibinfo  {journal} {Physical Review X}\ }\textbf {\bibinfo {volume} {14}},\ \bibinfo {pages} {021020} (\bibinfo {year} {2024})}\BibitemShut {NoStop}%
\bibitem [{\citenamefont {Rahmani}\ \emph {et~al.}(2018)\citenamefont {Rahmani}, \citenamefont {Loterie}, \citenamefont {Konstantinou}, \citenamefont {Psaltis},\ and\ \citenamefont {Moser}}]{rahmani2018multimode}%
  \BibitemOpen
  \bibfield  {author} {\bibinfo {author} {\bibfnamefont {B.}~\bibnamefont {Rahmani}}, \bibinfo {author} {\bibfnamefont {D.}~\bibnamefont {Loterie}}, \bibinfo {author} {\bibfnamefont {G.}~\bibnamefont {Konstantinou}}, \bibinfo {author} {\bibfnamefont {D.}~\bibnamefont {Psaltis}},\ and\ \bibinfo {author} {\bibfnamefont {C.}~\bibnamefont {Moser}},\ }\bibfield  {title} {\bibinfo {title} {Multimode optical fiber transmission with a deep learning network},\ }\href@noop {} {\bibfield  {journal} {\bibinfo  {journal} {Light: science \& applications}\ }\textbf {\bibinfo {volume} {7}},\ \bibinfo {pages} {69} (\bibinfo {year} {2018})}\BibitemShut {NoStop}%
\bibitem [{\citenamefont {Te{\u{g}}in}\ \emph {et~al.}(2020)\citenamefont {Te{\u{g}}in}, \citenamefont {Rahmani}, \citenamefont {Kakkava}, \citenamefont {Borhani}, \citenamefont {Moser},\ and\ \citenamefont {Psaltis}}]{teugin2020controlling}%
  \BibitemOpen
  \bibfield  {author} {\bibinfo {author} {\bibfnamefont {U.}~\bibnamefont {Te{\u{g}}in}}, \bibinfo {author} {\bibfnamefont {B.}~\bibnamefont {Rahmani}}, \bibinfo {author} {\bibfnamefont {E.}~\bibnamefont {Kakkava}}, \bibinfo {author} {\bibfnamefont {N.}~\bibnamefont {Borhani}}, \bibinfo {author} {\bibfnamefont {C.}~\bibnamefont {Moser}},\ and\ \bibinfo {author} {\bibfnamefont {D.}~\bibnamefont {Psaltis}},\ }\bibfield  {title} {\bibinfo {title} {Controlling spatiotemporal nonlinearities in multimode fibers with deep neural networks},\ }\href@noop {} {\bibfield  {journal} {\bibinfo  {journal} {Apl Photonics}\ }\textbf {\bibinfo {volume} {5}} (\bibinfo {year} {2020})}\BibitemShut {NoStop}%
\bibitem [{\citenamefont {Tzang}\ \emph {et~al.}(2018)\citenamefont {Tzang}, \citenamefont {Caravaca-Aguirre}, \citenamefont {Wagner},\ and\ \citenamefont {Piestun}}]{tzang2018adaptive}%
  \BibitemOpen
  \bibfield  {author} {\bibinfo {author} {\bibfnamefont {O.}~\bibnamefont {Tzang}}, \bibinfo {author} {\bibfnamefont {A.~M.}\ \bibnamefont {Caravaca-Aguirre}}, \bibinfo {author} {\bibfnamefont {K.}~\bibnamefont {Wagner}},\ and\ \bibinfo {author} {\bibfnamefont {R.}~\bibnamefont {Piestun}},\ }\bibfield  {title} {\bibinfo {title} {Adaptive wavefront shaping for controlling nonlinear multimode interactions in optical fibres},\ }\href@noop {} {\bibfield  {journal} {\bibinfo  {journal} {Nature Photonics}\ }\textbf {\bibinfo {volume} {12}},\ \bibinfo {pages} {368} (\bibinfo {year} {2018})}\BibitemShut {NoStop}%
\bibitem [{\citenamefont {Finot}\ \emph {et~al.}(2024)\citenamefont {Finot}, \citenamefont {Boscolo}, \citenamefont {Peng}, \citenamefont {Ermolaev}, \citenamefont {Sheveleva},\ and\ \citenamefont {Dudley}}]{finot2024machine}%
  \BibitemOpen
  \bibfield  {author} {\bibinfo {author} {\bibfnamefont {C.}~\bibnamefont {Finot}}, \bibinfo {author} {\bibfnamefont {S.}~\bibnamefont {Boscolo}}, \bibinfo {author} {\bibfnamefont {J.}~\bibnamefont {Peng}}, \bibinfo {author} {\bibfnamefont {A.}~\bibnamefont {Ermolaev}}, \bibinfo {author} {\bibfnamefont {A.}~\bibnamefont {Sheveleva}},\ and\ \bibinfo {author} {\bibfnamefont {J.~M.}\ \bibnamefont {Dudley}},\ }\bibfield  {title} {\bibinfo {title} {Machine learning for ultrafast nonlinear fibre photonics},\ }in\ \href@noop {} {\emph {\bibinfo {booktitle} {2024 24th International Conference on Transparent Optical Networks (ICTON)}}}\ (\bibinfo {organization} {IEEE},\ \bibinfo {year} {2024})\ pp.\ \bibinfo {pages} {1--4}\BibitemShut {NoStop}%
\bibitem [{\citenamefont {Reimer}\ \emph {et~al.}(2016)\citenamefont {Reimer}, \citenamefont {Kues}, \citenamefont {Roztocki}, \citenamefont {Wetzel}, \citenamefont {Grazioso}, \citenamefont {Little}, \citenamefont {Chu}, \citenamefont {Johnston}, \citenamefont {Bromberg}, \citenamefont {Caspani} \emph {et~al.}}]{reimer2016generation}%
  \BibitemOpen
  \bibfield  {author} {\bibinfo {author} {\bibfnamefont {C.}~\bibnamefont {Reimer}}, \bibinfo {author} {\bibfnamefont {M.}~\bibnamefont {Kues}}, \bibinfo {author} {\bibfnamefont {P.}~\bibnamefont {Roztocki}}, \bibinfo {author} {\bibfnamefont {B.}~\bibnamefont {Wetzel}}, \bibinfo {author} {\bibfnamefont {F.}~\bibnamefont {Grazioso}}, \bibinfo {author} {\bibfnamefont {B.~E.}\ \bibnamefont {Little}}, \bibinfo {author} {\bibfnamefont {S.~T.}\ \bibnamefont {Chu}}, \bibinfo {author} {\bibfnamefont {T.}~\bibnamefont {Johnston}}, \bibinfo {author} {\bibfnamefont {Y.}~\bibnamefont {Bromberg}}, \bibinfo {author} {\bibfnamefont {L.}~\bibnamefont {Caspani}}, \emph {et~al.},\ }\bibfield  {title} {\bibinfo {title} {Generation of multiphoton entangled quantum states by means of integrated frequency combs},\ }\href@noop {} {\bibfield  {journal} {\bibinfo  {journal} {Science}\ }\textbf {\bibinfo {volume} {351}},\ \bibinfo {pages} {1176} (\bibinfo {year} {2016})}\BibitemShut {NoStop}%
\bibitem [{\citenamefont {Uddin}\ \emph {et~al.}(2023)\citenamefont {Uddin}, \citenamefont {Rivera}, \citenamefont {Seyler}, \citenamefont {Salamin}, \citenamefont {Sloan}, \citenamefont {Roques-Carmes}, \citenamefont {Xu}, \citenamefont {Sander},\ and\ \citenamefont {Soljacic}}]{uddin2023ab}%
  \BibitemOpen
  \bibfield  {author} {\bibinfo {author} {\bibfnamefont {S.~Z.}\ \bibnamefont {Uddin}}, \bibinfo {author} {\bibfnamefont {N.}~\bibnamefont {Rivera}}, \bibinfo {author} {\bibfnamefont {D.}~\bibnamefont {Seyler}}, \bibinfo {author} {\bibfnamefont {Y.}~\bibnamefont {Salamin}}, \bibinfo {author} {\bibfnamefont {J.}~\bibnamefont {Sloan}}, \bibinfo {author} {\bibfnamefont {C.}~\bibnamefont {Roques-Carmes}}, \bibinfo {author} {\bibfnamefont {S.}~\bibnamefont {Xu}}, \bibinfo {author} {\bibfnamefont {M.}~\bibnamefont {Sander}},\ and\ \bibinfo {author} {\bibfnamefont {M.}~\bibnamefont {Soljacic}},\ }\bibfield  {title} {\bibinfo {title} {An ab initio framework for understanding and controlling quantum fluctuations in highly multimoded light-matter systems},\ }\href@noop {} {\bibfield  {journal} {\bibinfo  {journal} {arXiv preprint arXiv:2311.05535}\ } (\bibinfo {year} {2023})}\BibitemShut {NoStop}%
\bibitem [{\citenamefont {Pontula}\ \emph {et~al.}(2024)\citenamefont {Pontula}, \citenamefont {Salamin}, \citenamefont {Roques-Carmes},\ and\ \citenamefont {Solja{\v{c}}i{\'c}}}]{pontula2024shaping}%
  \BibitemOpen
  \bibfield  {author} {\bibinfo {author} {\bibfnamefont {S.}~\bibnamefont {Pontula}}, \bibinfo {author} {\bibfnamefont {Y.}~\bibnamefont {Salamin}}, \bibinfo {author} {\bibfnamefont {C.}~\bibnamefont {Roques-Carmes}},\ and\ \bibinfo {author} {\bibfnamefont {M.}~\bibnamefont {Solja{\v{c}}i{\'c}}},\ }\bibfield  {title} {\bibinfo {title} {Shaping quantum noise through cascaded nonlinear processes in a dissipation-engineered multimode cavity},\ }\href@noop {} {\bibfield  {journal} {\bibinfo  {journal} {PRX Quantum}\ }\textbf {\bibinfo {volume} {5}},\ \bibinfo {pages} {040345} (\bibinfo {year} {2024})}\BibitemShut {NoStop}%
\bibitem [{\citenamefont {Goda}\ \emph {et~al.}(2009)\citenamefont {Goda}, \citenamefont {Solli}, \citenamefont {Tsia},\ and\ \citenamefont {Jalali}}]{goda2009theory}%
  \BibitemOpen
  \bibfield  {author} {\bibinfo {author} {\bibfnamefont {K.}~\bibnamefont {Goda}}, \bibinfo {author} {\bibfnamefont {D.~R.}\ \bibnamefont {Solli}}, \bibinfo {author} {\bibfnamefont {K.~K.}\ \bibnamefont {Tsia}},\ and\ \bibinfo {author} {\bibfnamefont {B.}~\bibnamefont {Jalali}},\ }\bibfield  {title} {\bibinfo {title} {Theory of amplified dispersive fourier transformation},\ }\href@noop {} {\bibfield  {journal} {\bibinfo  {journal} {Physical Review A—Atomic, Molecular, and Optical Physics}\ }\textbf {\bibinfo {volume} {80}},\ \bibinfo {pages} {043821} (\bibinfo {year} {2009})}\BibitemShut {NoStop}%
\bibitem [{\citenamefont {Goda}\ and\ \citenamefont {Jalali}(2013)}]{goda2013dispersive}%
  \BibitemOpen
  \bibfield  {author} {\bibinfo {author} {\bibfnamefont {K.}~\bibnamefont {Goda}}\ and\ \bibinfo {author} {\bibfnamefont {B.}~\bibnamefont {Jalali}},\ }\bibfield  {title} {\bibinfo {title} {Dispersive fourier transformation for fast continuous single-shot measurements},\ }\href@noop {} {\bibfield  {journal} {\bibinfo  {journal} {Nature Photonics}\ }\textbf {\bibinfo {volume} {7}},\ \bibinfo {pages} {102} (\bibinfo {year} {2013})}\BibitemShut {NoStop}%
\bibitem [{\citenamefont {Godin}\ \emph {et~al.}(2022)\citenamefont {Godin}, \citenamefont {Sader}, \citenamefont {Khodadad~Kashi}, \citenamefont {Hanzard}, \citenamefont {Hideur}, \citenamefont {Moss}, \citenamefont {Morandotti}, \citenamefont {Genty}, \citenamefont {Dudley}, \citenamefont {Pasquazi} \emph {et~al.}}]{godin2022recent}%
  \BibitemOpen
  \bibfield  {author} {\bibinfo {author} {\bibfnamefont {T.}~\bibnamefont {Godin}}, \bibinfo {author} {\bibfnamefont {L.}~\bibnamefont {Sader}}, \bibinfo {author} {\bibfnamefont {A.}~\bibnamefont {Khodadad~Kashi}}, \bibinfo {author} {\bibfnamefont {P.-H.}\ \bibnamefont {Hanzard}}, \bibinfo {author} {\bibfnamefont {A.}~\bibnamefont {Hideur}}, \bibinfo {author} {\bibfnamefont {D.~J.}\ \bibnamefont {Moss}}, \bibinfo {author} {\bibfnamefont {R.}~\bibnamefont {Morandotti}}, \bibinfo {author} {\bibfnamefont {G.}~\bibnamefont {Genty}}, \bibinfo {author} {\bibfnamefont {J.~M.}\ \bibnamefont {Dudley}}, \bibinfo {author} {\bibfnamefont {A.}~\bibnamefont {Pasquazi}}, \emph {et~al.},\ }\bibfield  {title} {\bibinfo {title} {Recent advances on time-stretch dispersive fourier transform and its applications},\ }\href@noop {} {\bibfield  {journal} {\bibinfo  {journal} {Advances in Physics: X}\ }\textbf {\bibinfo {volume} {7}},\ \bibinfo {pages} {2067487} (\bibinfo {year} {2022})}\BibitemShut {NoStop}%
\bibitem [{\citenamefont {Wang}\ \emph {et~al.}(2020)\citenamefont {Wang}, \citenamefont {Wang}, \citenamefont {Zhang}, \citenamefont {Guo}, \citenamefont {Ma}, \citenamefont {Huang}, \citenamefont {Song}, \citenamefont {Ge}, \citenamefont {Liu},\ and\ \citenamefont {Zhang}}]{wang2020recent}%
  \BibitemOpen
  \bibfield  {author} {\bibinfo {author} {\bibfnamefont {Y.}~\bibnamefont {Wang}}, \bibinfo {author} {\bibfnamefont {C.}~\bibnamefont {Wang}}, \bibinfo {author} {\bibfnamefont {F.}~\bibnamefont {Zhang}}, \bibinfo {author} {\bibfnamefont {J.}~\bibnamefont {Guo}}, \bibinfo {author} {\bibfnamefont {C.}~\bibnamefont {Ma}}, \bibinfo {author} {\bibfnamefont {W.}~\bibnamefont {Huang}}, \bibinfo {author} {\bibfnamefont {Y.}~\bibnamefont {Song}}, \bibinfo {author} {\bibfnamefont {Y.}~\bibnamefont {Ge}}, \bibinfo {author} {\bibfnamefont {J.}~\bibnamefont {Liu}},\ and\ \bibinfo {author} {\bibfnamefont {H.}~\bibnamefont {Zhang}},\ }\bibfield  {title} {\bibinfo {title} {Recent advances in real-time spectrum measurement of soliton dynamics by dispersive fourier transformation},\ }\href@noop {} {\bibfield  {journal} {\bibinfo  {journal} {Reports on Progress in Physics}\ }\textbf {\bibinfo {volume} {83}},\ \bibinfo {pages} {116401} (\bibinfo {year} {2020})}\BibitemShut {NoStop}%
\bibitem [{\citenamefont {Lapre}\ \emph {et~al.}(2020)\citenamefont {Lapre}, \citenamefont {Billet}, \citenamefont {Meng}, \citenamefont {Genty},\ and\ \citenamefont {Dudley}}]{lapre2020dispersive}%
  \BibitemOpen
  \bibfield  {author} {\bibinfo {author} {\bibfnamefont {C.}~\bibnamefont {Lapre}}, \bibinfo {author} {\bibfnamefont {C.}~\bibnamefont {Billet}}, \bibinfo {author} {\bibfnamefont {F.}~\bibnamefont {Meng}}, \bibinfo {author} {\bibfnamefont {G.}~\bibnamefont {Genty}},\ and\ \bibinfo {author} {\bibfnamefont {J.~M.}\ \bibnamefont {Dudley}},\ }\bibfield  {title} {\bibinfo {title} {Dispersive fourier transform characterization of multipulse dissipative soliton complexes in a mode-locked soliton-similariton laser},\ }\href@noop {} {\bibfield  {journal} {\bibinfo  {journal} {OSA Continuum}\ }\textbf {\bibinfo {volume} {3}},\ \bibinfo {pages} {275} (\bibinfo {year} {2020})}\BibitemShut {NoStop}%
\bibitem [{\citenamefont {Wetzel}\ \emph {et~al.}(2012)\citenamefont {Wetzel}, \citenamefont {Stefani}, \citenamefont {Larger}, \citenamefont {Lacourt}, \citenamefont {Merolla}, \citenamefont {Sylvestre}, \citenamefont {Kudlinski}, \citenamefont {Mussot}, \citenamefont {Genty}, \citenamefont {Dias} \emph {et~al.}}]{wetzel2012real}%
  \BibitemOpen
  \bibfield  {author} {\bibinfo {author} {\bibfnamefont {B.}~\bibnamefont {Wetzel}}, \bibinfo {author} {\bibfnamefont {A.}~\bibnamefont {Stefani}}, \bibinfo {author} {\bibfnamefont {L.}~\bibnamefont {Larger}}, \bibinfo {author} {\bibfnamefont {P.-A.}\ \bibnamefont {Lacourt}}, \bibinfo {author} {\bibfnamefont {J.-M.}\ \bibnamefont {Merolla}}, \bibinfo {author} {\bibfnamefont {T.}~\bibnamefont {Sylvestre}}, \bibinfo {author} {\bibfnamefont {A.}~\bibnamefont {Kudlinski}}, \bibinfo {author} {\bibfnamefont {A.}~\bibnamefont {Mussot}}, \bibinfo {author} {\bibfnamefont {G.}~\bibnamefont {Genty}}, \bibinfo {author} {\bibfnamefont {F.}~\bibnamefont {Dias}}, \emph {et~al.},\ }\bibfield  {title} {\bibinfo {title} {Real-time full bandwidth measurement of spectral noise in supercontinuum generation},\ }\href@noop {} {\bibfield  {journal} {\bibinfo  {journal} {Scientific reports}\ }\textbf {\bibinfo {volume} {2}},\ \bibinfo {pages} {882} (\bibinfo {year} {2012})}\BibitemShut {NoStop}%
\bibitem [{\citenamefont {Godin}\ \emph {et~al.}(2013)\citenamefont {Godin}, \citenamefont {Wetzel}, \citenamefont {Sylvestre}, \citenamefont {Larger}, \citenamefont {Kudlinski}, \citenamefont {Mussot}, \citenamefont {Salem}, \citenamefont {Zghal}, \citenamefont {Genty}, \citenamefont {Dias} \emph {et~al.}}]{godin2013real}%
  \BibitemOpen
  \bibfield  {author} {\bibinfo {author} {\bibfnamefont {T.}~\bibnamefont {Godin}}, \bibinfo {author} {\bibfnamefont {B.}~\bibnamefont {Wetzel}}, \bibinfo {author} {\bibfnamefont {T.}~\bibnamefont {Sylvestre}}, \bibinfo {author} {\bibfnamefont {L.}~\bibnamefont {Larger}}, \bibinfo {author} {\bibfnamefont {A.}~\bibnamefont {Kudlinski}}, \bibinfo {author} {\bibfnamefont {A.}~\bibnamefont {Mussot}}, \bibinfo {author} {\bibfnamefont {A.~B.}\ \bibnamefont {Salem}}, \bibinfo {author} {\bibfnamefont {M.}~\bibnamefont {Zghal}}, \bibinfo {author} {\bibfnamefont {G.}~\bibnamefont {Genty}}, \bibinfo {author} {\bibfnamefont {F.}~\bibnamefont {Dias}}, \emph {et~al.},\ }\bibfield  {title} {\bibinfo {title} {Real time noise and wavelength correlations in octave-spanning supercontinuum generation},\ }\href@noop {} {\bibfield  {journal} {\bibinfo  {journal} {Optics Express}\ }\textbf {\bibinfo {volume} {21}},\ \bibinfo {pages} {18452} (\bibinfo {year} {2013})}\BibitemShut {NoStop}%
\bibitem [{\citenamefont {Dudley}\ \emph {et~al.}(2006)\citenamefont {Dudley}, \citenamefont {Genty},\ and\ \citenamefont {Coen}}]{dudley2006supercontinuum}%
  \BibitemOpen
  \bibfield  {author} {\bibinfo {author} {\bibfnamefont {J.~M.}\ \bibnamefont {Dudley}}, \bibinfo {author} {\bibfnamefont {G.}~\bibnamefont {Genty}},\ and\ \bibinfo {author} {\bibfnamefont {S.}~\bibnamefont {Coen}},\ }\bibfield  {title} {\bibinfo {title} {Supercontinuum generation in photonic crystal fiber},\ }\href@noop {} {\bibfield  {journal} {\bibinfo  {journal} {Reviews of modern physics}\ }\textbf {\bibinfo {volume} {78}},\ \bibinfo {pages} {1135} (\bibinfo {year} {2006})}\BibitemShut {NoStop}%
\bibitem [{\citenamefont {Dudley}\ and\ \citenamefont {Taylor}(2010)}]{dudley2010supercontinuum}%
  \BibitemOpen
  \bibfield  {author} {\bibinfo {author} {\bibfnamefont {J.~M.}\ \bibnamefont {Dudley}}\ and\ \bibinfo {author} {\bibfnamefont {J.~R.}\ \bibnamefont {Taylor}},\ }\href@noop {} {\emph {\bibinfo {title} {Supercontinuum generation in optical fibers}}}\ (\bibinfo  {publisher} {Cambridge university press},\ \bibinfo {year} {2010})\BibitemShut {NoStop}%
\bibitem [{\citenamefont {Drummond}\ and\ \citenamefont {Corney}(2001)}]{drummond2001quantum}%
  \BibitemOpen
  \bibfield  {author} {\bibinfo {author} {\bibfnamefont {P.~D.}\ \bibnamefont {Drummond}}\ and\ \bibinfo {author} {\bibfnamefont {J.~F.}\ \bibnamefont {Corney}},\ }\bibfield  {title} {\bibinfo {title} {Quantum noise in optical fibers. i. stochastic equations},\ }\href@noop {} {\bibfield  {journal} {\bibinfo  {journal} {JOSA B}\ }\textbf {\bibinfo {volume} {18}},\ \bibinfo {pages} {139} (\bibinfo {year} {2001})}\BibitemShut {NoStop}%
\bibitem [{\citenamefont {Blow}\ and\ \citenamefont {Wood}(1989)}]{blow1989theoretical}%
  \BibitemOpen
  \bibfield  {author} {\bibinfo {author} {\bibfnamefont {K.~J.}\ \bibnamefont {Blow}}\ and\ \bibinfo {author} {\bibfnamefont {D.}~\bibnamefont {Wood}},\ }\bibfield  {title} {\bibinfo {title} {Theoretical description of transient stimulated raman scattering in optical fibers},\ }\href@noop {} {\bibfield  {journal} {\bibinfo  {journal} {IEEE Journal of Quantum Electronics}\ }\textbf {\bibinfo {volume} {25}},\ \bibinfo {pages} {2665} (\bibinfo {year} {1989})}\BibitemShut {NoStop}%
\end{thebibliography}%

\end{document}

\rmfamily

\title{Supplementary Information for: Probing intensity noise in ultrafast pulses using the dispersive Fourier transform augmented by quantum sensitivity analysis}

%-----AUTHORS AND AFFILIATIONS-----
\author{Shiekh Zia Uddin$^{1,2,\dagger}$}
%\email{suddin@mit.edu}
\author{Sahil Pontula$^{1,2,3,\dagger}$}
%\email{spontula@mit.edu}
\author{Jiaxin Liu$^4$}
\author{Shutao Xu$^5$}
\author{Seou Choi$^{2,3}$}
\author{Michelle Y. Sander$^5$}
\author{Marin Solja\v{c}i\'{c}$^{1,3}$}
\affiliation{$^1$ Department of Physics, MIT, Cambridge, MA 02139, USA. \\
$^2$ Research Laboratory of Electronics, MIT, Cambridge, MA 02139,  USA.\\
$^3$ Department of Electrical Engineering and Computer Science, MIT 02139, Cambridge, MA, USA.\\
$^4$ Department of Physics, Imperial College London, South Kensington, London SW7 2BW, UK.\\
$^5$ Department of Electrical and Computer Engineering and BU Photonics Center, Boston University, Boston, MA 02215, USA.\\
$\dagger$ Denotes equal contribution. Email: suddin@mit.edu, spontula@mit.edu}
\noindent	

\noindent
%\tableofcontents

\clearpage

\renewcommand{\sp}{\sigma_+}
\newcommand{\sm}{\sigma_-}

%-----CHANGE SETUP FOR PARAGRAPH INDENTS AND SKIPS-----
\setlength{\parindent}{0em}
\setlength{\parskip}{.5em}
\vspace*{-2em}

%-------------------------------------
%------------- MAIN TEXT -------------
%-------------------------------------

\newcommand{\bin}{b_{\text{in}}}
\newcommand{\bbarin}{\bar{b}_{\text{in}}}
%More lasing oriented stuff
\begin{abstract}
    In this Supplementary Information (S.I.), we present results figures complementing discussions in the main text as well as a discussion of the main steps necessary for post-processing data collected by the dispersive Fourier transform in our experiments.
    % additional details on computational methods as well as 
\end{abstract}

\maketitle

%\tableofcontents

\newpage

%\section{Generalized nonlinear Schrodinger equation and quantum sensitivity analysis}

%\section{Propagation length dependence}

%\section{Fine power dependence}

\section{Post-processing steps for dispersive Fourier transform data}

Although the collection of the dispersive Fourier transform (DFT) data is relatively straightforward, a few caveats should be noted with respect to post-processing steps. Since the sampling rate of the oscilloscope is not in general an integer multiple of the pump laser's repetition rate, the oscilloscope traces need to be resampled and sliced into signals corresponding to individual input pulses. One can then use the average intensity measurement provided by an optical spectrum analyzer (OSA) as a reference to convert these sliced traces to spectra in the frequency domain. In order to do so, two considerations are necessary. First, the start of each oscilloscope measurement does not coincide with the arrival of a stretched pulse at the photodiode, so the positions where each trace is sliced need to be shifted together by some offset. Second, the conversion factor between the time and frequency domains must be computed. In our work, both the offset and the conversion factor were found by iteratively applying a resampling factor to the mean waveform of the sliced traces and shifting the start of waveform, while maximizing the overlap between each OSA spectrum and the mean oscilloscope waveform (at a fixed power). The optimization was performed by applying a single conversion factor at all input powers and manually adjusting the offsets at each power to maximize agreement with the OSA spectra at all powers. With the optimized conversion factor for all powers and offsets specific to the measurements at each power, the sliced oscilloscope traces could then be converted to shot-by-shot spectra in the frequency domain.

\section{Supplementary figures}

\begin{figure}[t]
    \centering
    \includegraphics[scale=0.6]{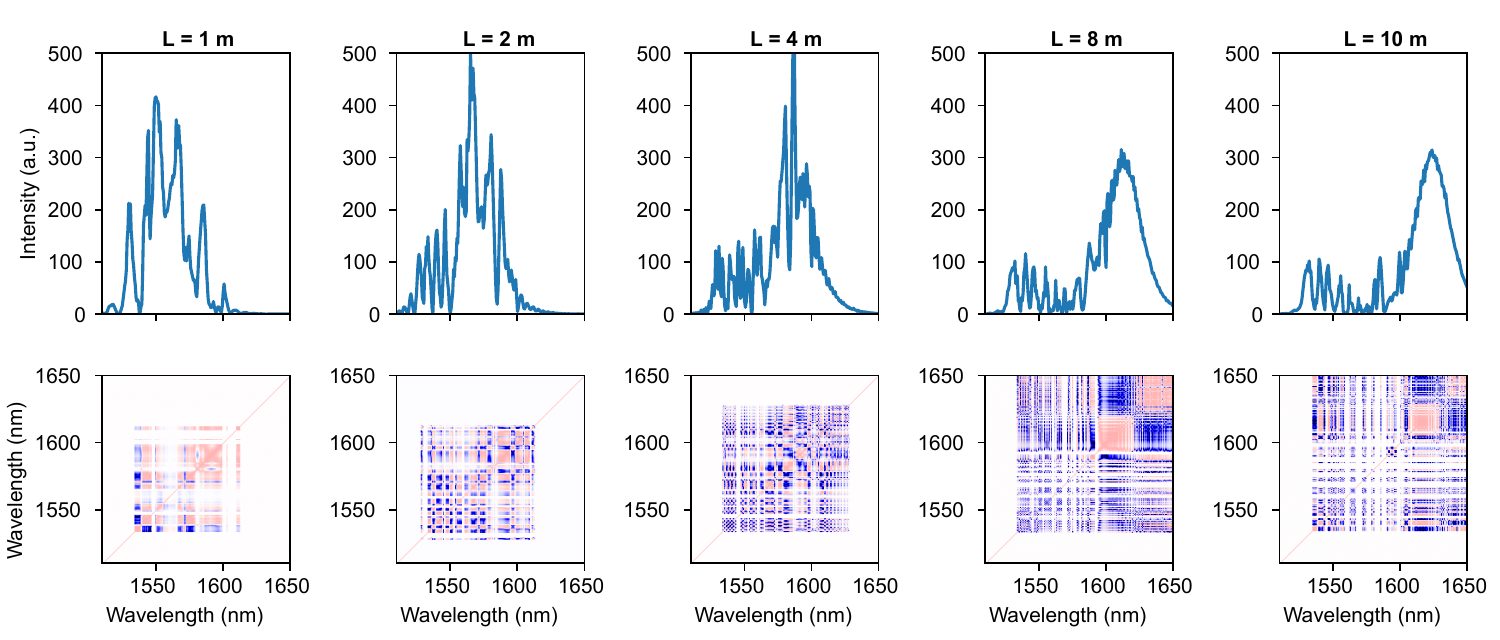}
    \caption{\textbf{Dependence of mean field spectra and wavelength pair correlations on propagation length.} Top row: mean field spectra calculated for different propagation lengths using GNLSE with parameters provided in main text at input power 120 mW. Bottom row: corresponding sum noise of wavelength pairs for each propagation length. Loosely speaking, changing the fiber length serves as a proxy for changing the input power of the pulse in this scenario, with longer fiber lengths shifting the Raman soliton to longer wavelengths.}
    \label{fig:f1}
\end{figure}

\afterpage{
\begin{figure}[t]
    \centering
    \includegraphics[scale=0.7]{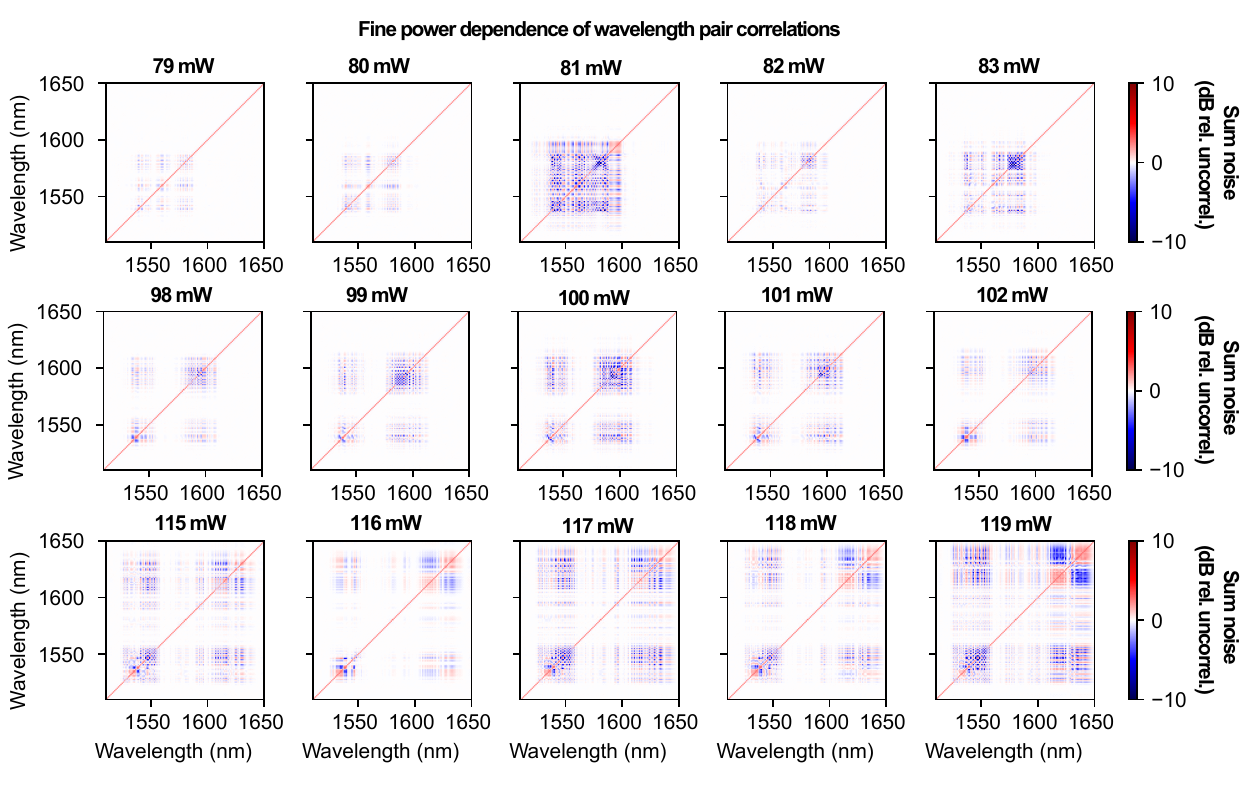}
    \caption{\textbf{Fine power dependence of wavelength pair correlations.} Correlation heatmaps at finer input power resolution for three different power regimes after soliton fission. The sensitive power dependence of certain noise features may be attributed to the sensitive Kerr nonlinear phase shifts at different wavelengths.}
    \label{fig:f2}
\end{figure}
}

%\bibliographystyle{unsrt}
%\bibliography{supp.bib}